# ESCAPE

Preparing Forecasting Systems for the Next generation of Supercomputers

# D2.5 Report on the performance portability demonstrated for the relevant Weather & Climate Dwarfs

Dissemination Level: Public

This project has received funding from the European Union's Horizon 2020 research and innovation programme under grant agreement No 67162

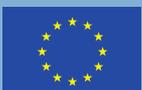

Funded by the European Union

Co-ordinated by

# ESCAPE

**Energy-efficient Scalable Algorithms for Weather Prediction at Exascale**

Author **Carlos Osuna**

Date **06/06/2018**

# Table of Contents

# Figures

## Tables

# 1 Executive Summary

This deliverable provides an evaluation of the work performed within ESCAPE to port different dwarfs to accelerators, using different programming models. A key metric of the evaluation is the performance portability of the resulting porting efforts. Portability means that a single source code containing the numerical operators can be compiled and run in multiple architectures, while performance portability additionally requires that the single source code runs efficiently in all the different architectures.

As results of other deliverables like D2.1, D2.4 ESCAPE provides a collection of dwarfs ported to different computing architectures like traditional CPUs, Intel XeonPhi and NVIDIA GPUs. Additionally D3.3 went through an optimization process to obtain efficient and energy efficient dwarfs.

In this deliverable we present a review of the different programming models employed and their use to port various dwarfs of ESCAPE. A final evaluation of the different approaches based on different metrics like performance portability, readability of the numerical methods, efforts to port a dwarf and efficiency of the implementation obtained is reported.

# 2 Introduction

## 2.1 Scope of this deliverable

### 2.1.1 Objectives of this deliverable

The goal of this deliverable is to establish a comparison between the different programming models used in the ESCAPE dwarfs in terms of usability, portability and performance obtained based on the work performed by the ESCAPE project porting the different dwarfs to accelerators and modern architectures.

### 2.1.2 Work performed in this deliverable

In this deliverable we evaluate the different programming models utilized to port the various dwarfs of ESCAPE to accelerators. First we present a review of the DSL approach, implemented by the deliverable D2.4. The use of other programming models like OpenMP and OpenACC was reported in deliverables D2.1 and D3.3. Finally we establish a comparison and evaluation of the different approaches based on concrete examples of the MPDATA dwarf.

### 2.1.3 Deviations and counter measures

There are no deviations with respect to the planned work.

# 3 Review of programming models for portability of Weather and Climate models

## 3.1 GridTools DSL for structured lat-lon grids

### 3.1.1 GridTools overview

The GridTools framework provides a set of tools for developing numerical methods of weather and climate applications. The emergence of new computing architectures and accelerators in the supercomputing systems where weather and climate applications are run are posing a challenge to efficiently maintain and run weather models. Typically weather models are complex systems with large codebases (from hundred thousand to million lines of code). Adapting models to new computing architectures is a daunting task for multiple reasons: computing architectures are quite different among them. Often they offer different memory spaces that must be managed explicitly (like GPUs), efficient computations on gridded fields require storing the multidimensional fields with different memory layouts, etc. And the nested loops over dimensions and performance optimizations (such as tiling/loop blocking, loop fusion, etc.) are specific to each hardware architecture. Additionally they might require the use of different programming models.

The main goal of the GridTools library is to provide abstractions for developing weather and climate models that perform efficiently on multiple architectures. As shown in Figure 1, it provides the following library components:

1. Multidimensional storages. It provides the functionality for storage and efficient access of multidimensional fields with full control on the data layout, alignment and padding.
2. Halo exchanges and boundary conditions. It allows implementing halo exchanges and general boundary conditions (in multiple dimensions) that abstract away the architecture dependent implementation.
3. Stencil DSL. It provides a domain specific language that is used to describe the mathematical operations of weather and climate code abstracting the entire details specific to efficient implementation for each computing architecture, like looping, nested tiling, software managed caches, etc.
4. Computing architecture backend. The specification of a stencil DSL is translated into efficient code for a specific architecture by the architecture backend. Currently it supports backends for traditional CPUs, XeonPhi and NVIDIA GPUs. The CPU and XeonPhi backends are implemented using OpenMP while the GPU backends uses the CUDA programming model.

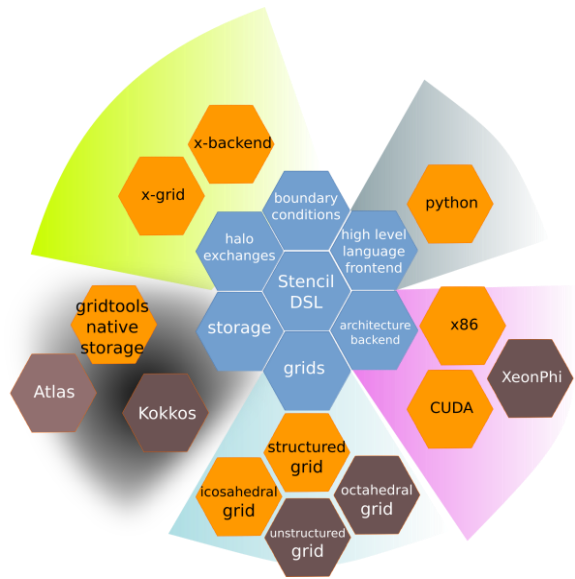

*Figure 1 Components and structure of the GridTools library*

### 3.1.2 DSL syntax

In this section we will describe the syntax of the main stencil DSL component which allows developing numerical methods in a concise manner for the typical computational patterns of weather and climate applications.

A GridTools stencil definition is made up of two main components: 1) an **operator syntax** that describes, using the DSL language, the grid point operation applied to one or more fields and 2) DSL language elements that allow for **composition of multiple operators** into a single kernel.

The first component allows to describe equations in a concise manner, removing all the details of an implementation that are specific to each architecture, like the use of a parallel programming model, loop nesting and loop fusion, or efficient use of the memory subsystem of each computing architecture.

An example of a Laplace operator on lat-lon grids is shown in Listing 1.

```
1  struct Laplace {
2    using in = accessor<0, in, extent<-1,1,-1,1>>;
3    using out = accessor<1, out, extent<>>;
4
5    template<typename Evaluation>
6    GT_FUNCTION static void Do(Evaluation& eval) {
7      eval(lap()) = eval(-4*u()+u(i+1)+u(i-1)+u(j-1)+u(j+1));
8    }
9  };
```

*Listing 1 Example of a Laplace operator implemented using the GridTools DSL syntax*

In order to access neighbor grid elements, there are multiple syntax elements. The following examples have equivalent meaning.

```
1        eval(lap()) = eval(-4*u()+u(i+1)+u(i-1)+u(j-1)+u(j+1));
2        eval(lap()) = eval(-4*u+u{1,0,0}+u{-1,0,0}+u{0,-1,0}+u{0,1,0});
3        eval(lap()) = eval(-4*u+u(1,0)+u(-1,0)+u(0,-1)+u(0,1));
```

The second component allows combining multiple of the basic operators into a single kernel. Due to the strong data locality of operations on fields present in weather and climate codes, the assembly of multiple operators in a single kernel is used by the library to increase data locality of the algorithms and make efficient use of the different levels of cache memory present in each architecture.

An example of a horizontal diffusion kernel composition is shown in Listing 2.

```
1 auto horizontal_diffusion = make_computation<backend<Cuda, structured > >(
2   grid, p_in() = u, p_out() = u_diff, p_coeff() = coeff,
3   make_multistage(execute<parallel>,
4       define_caches(cache<IJ, local >(p_lap(), p_flx(), p_fly())),
5       make_stage<lap_function>(p_lap(), p_in()),
6       make_stage<flx_function>(p_flx(), p_in(), p_lap()),
7       make_stage<fly_function>(p_fly(), p_in(), p_lap()),
8       make_stage<div_function>(p_out(), p_in(), p_flx(), p_fly(),
p_coeff())));
```

*Listing 2 Stencil composition of multiple DSL operators to form a computation object*

Line 1 builds the computation object that will run the stencil computations on a set of gridded fields, where the *backend* keyword is used to select the target computing architecture (a NVIDIA GPU in this case).

Line 2 will bind the parameters to actual storage fields where the computation will run (in this case horizontal wind speed *u* field)

Syntax in lines 5-8 is used to combine multiple operators in a single kernel object and the *cache* syntax is used to describe data reuse patterns

### 3.1.3 Development of DSL for weather models on global grids

The main goal of the GridTools library is to provide a solution to portability and performance portability for weather models. However, the main production product of the GridTools focused on solutions for lat-lon grid models like COSMO. The work developed in ESCAPE aimed at extending the DSL support for irregular grids and efficient generation of backends for multiple architectures. In this section we summarize the results of the ESCAPE efforts to extend the library and in particular the DSL syntax to support methods on irregular grids.

### 3.1.4 Grids

In order to provide solutions for portability of the different dwarfs of the ESCAPE project, GridTools had to be extended and support a broad type of irregular grids, like the icosahedral (shown in Figure 2) or octahedral grids.

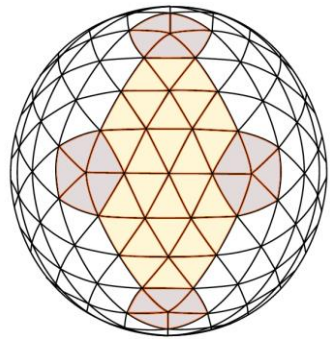

*Figure 2 Domain decomposition of the icosahedral grid. The figure highlights one of the main diamonds of the original icosahedron*

Although most of the dwarfs of the ESCAPE project are implemented on the octahedral grid, in this work we proposed a new DSL syntax that is valid for any type of grid, based on concepts present in all type of grids: field location and stencil operations over fields of different location types.

### 3.1.4.1 Grid Staggering

Staggered grids, like the Arakawa C grid are widely used in the global weather models. The advantage of staggered grids like the C grid is the effective higher spatial resolution for differential operators like the divergence and pressure gradient where $\Delta x$ is half as compared to A grids, where all the fields are located at the nodes of the mesh. In addition the energy propagation has the wrong direction in the A and B grids for waves with a scale close to the grid scale.

The A and C grids are shown for lat-lon type of grids in Figure 3. In a lat-lon C grid, horizontal wind speed velocity components u and v are located at the center of the horizontal and vertical edge (respectively) between two nodes.

Even if the spatial positions of u,v fields are *staggered*, models express those fields as arrays of values with the same index space, usually noted as (i,j,k), assuming the convention that u(i,j) is located $\Delta x / 2$ at the right of q(i,j) in the x direction and v(i,j) is $\Delta x / 2$ above q(i,j).

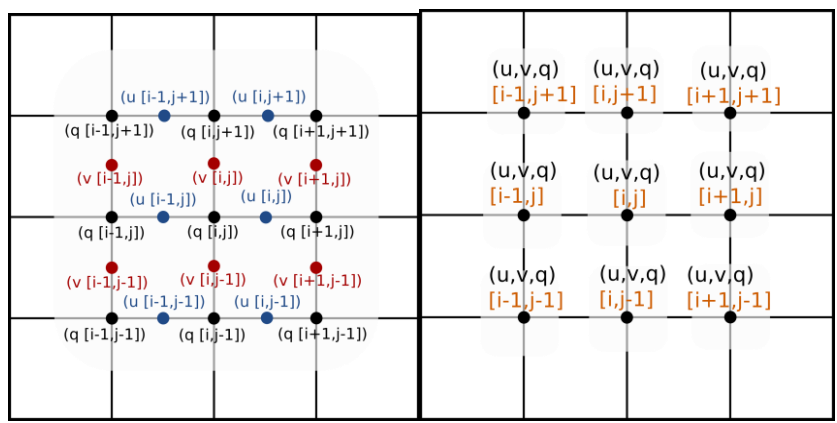

*Figure 3 Left: Indexing of different fields on a lat-lon A grid. Right: indexing of different fields on a lat-lon C grid*

### 3.1.4.2 Irregular Grids

An irregular (i.e. non Cartesian grid) is a general type of grid composed by simple shapes such as triangles, quadrilaterals or pentagons. Since the elementary shape is not anymore only quadrilaterals, the indexing scheme based on tuples of *(row index, column index)* employed by lat-lon models is not valid anymore. Furthermore, staggered grids like the C grid require an indexing of elements capable of locating fields on edges, nodes and cell centers.

The simple rules of the lat-lon grids to locate neighbor grid points of a field using i±1, j±1 offsets cannot be used in irregular staggered grids.

Figure 4 shows a typical placement of fields for models on a staggered C grid on triangles, like the octahedral or icosahedral C grid. Fields can be located at edge centers (velocities, gradients …), nodes (vorticity) or cell centers (temperature, pressure,…).

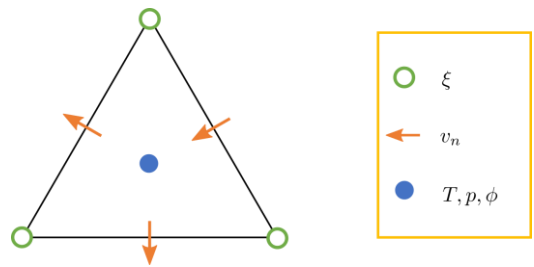

*Figure 4 Example of location of different type fields on a triangular C-grid*

.
In order to index fields in their corresponding location and to establish the connectivity with neighbors, the GridTools backend for irregular grids introduces the concept of a **location** type, which in general for any type of elementary shape, can be *edges, cell centers and vertices.* The location type can be used in the declaration of the storages that hold fields as well as in the operators.

Another concept of the GridTools backend for irregular grids is the **topology** of the grids, which knows the connectivity among the grid points in each location type for each of the supported grids.

Some irregular grids, like the octahedral/icosahedral grid will still retain a structure which allows deriving rules for extracting the connectivity of the topology without the use of unstructured meshes. Other grids, like a generic reduced Gaussian have a structure, although more complex and do not allow to derive easy methods to extract the connectivity. The last group will exhibit a totally irregular pattern and will require the use of an unstructured mesh.

The GridTools topology can be configured in two modes for structured and unstructured meshes.

The following snippet shows an example of constructing a topology for the (structured) octahedral grid and allocates fields in a specific location type

```
1 grid_topology<octahedral> octTopology(128,128,80);
2 auto rho = octTopology.make_storage<octahedral::vertices, double>("rho");
3 auto Vstar = octTopology.make_storage<octahedral::vertices, double,
      selector<1,1,1,1,1>>("Vstar", 3);
4 auto edge_length = octTopology.make_storage<octahedral::edges, double,
      selector<1,1,1,0>>("edge_length");
```

Line 1 constructs a grid topology for an octahedral (structured) type of grid.

Line 2 will allocate a field (density) with a location on *nodes* and a default number of dimensions 3 (with the domain sizes of the topology in line 1.

The structured backends of the irregular grid incorporate an extra dimension for the indexing of the grid points (color), so that 3D fields are represented as 4D storages. More details on the indexing and coloring of the grid will be explained in Section 3.2.

Line 3 is using an optional keyword to specify the requested dimensions. The four first dimensions represent (row, color, column, vertical layer). In this case Vstar is requested as a field with all default dimensions and an extra dimension of size 3 to host the three components of a velocity field located in nodes.

Line 4 is requesting a two dimensional field located on edges in order to store the edge length.

The allocated storages, which contain the concept of a location type for irregular grids, will be later passed to the stencil composition (see Listing 2 for example) that will build the computations of a particular model component. They can also be used, independently of the DSL, as C++ storages. The following example will initialize a multidimensional storage on cells, with an extra dimension to store a weight for each neighbor edge of a cell, using standard C++ code:

```cpp
1  auto  weight  =  octTopology.make_storage<octahedral::cells, double,
selector<1,1,1,0,1>>("weight", 3);
2 auto edge_length = octTopology.make_storage<octahedral::edges,
double, selector<1,1,1,0>>("edge_length");
3 auto cell_area = octTopology.make_storage<octahedral::cells, double,
selector<1,1,1,0>>("cell_area");
4   auto weight_view = make_host_view(weight);
5   auto edgelength_view = make_host_view(edge_length);
6   auto cellarea_view = make_host_view(cell_area);
7   for(auto i: range(0,isize)) {
8     for(auto cc: range(0, octahedral::vertices::ncolors)) {
9       for(auto j: range(0,isize)) {
10        for(auto ec: range(0, octahedral::edges::ncolors)) {
11          weight_view(i,cc,j,ec) = edgelength_view(i,ec,j) /
cellarea_view(i,cc,j);
12        }
13      }
14    }
15 }
```

*Listing 3 C++ example of a computation of a weight field using GridTools storages on different location types.*

Line 1-3 allocates the different fields as GridTools storage with a location type. In order to access the data, a multidimensional view of the storage is constructed in Lines 4-6. For computing architecture backends that require the management of data in different memory spaces like the GPUs, the GridTools storages allow to explicitly manage the data of each memory space and synchronize the data among the different memory spaces.

The construct *make_device_view* will provide a view to data allocated in the device (similar to the host views used in Listing 3).

Explicit synchronization methods allow to control the memory transfers between the device and the host views.

### 3.1.5 Stencil Operations

As shown in Section 3.1.2, the basic syntax of the lat-lon grid backend of the DSL for stencil operations is based on the specification of the relative offset of each field access, i.e. `eval(u(i+1))` or `eval(u{1,0,0})`. Both refer to the neighbor grid point of the *u* field in the positive *x* direction.

In irregular grids, the connectivity between grid points in different location types cannot be expressed anymore with this syntax for lat-lon grid operators.

In general, for irregular grids, there will be 9 different types of stencil operators (see Figure 5), depending on the location type of the fields where the stencil operates on, and the location type of the neighboring fields being accessed.

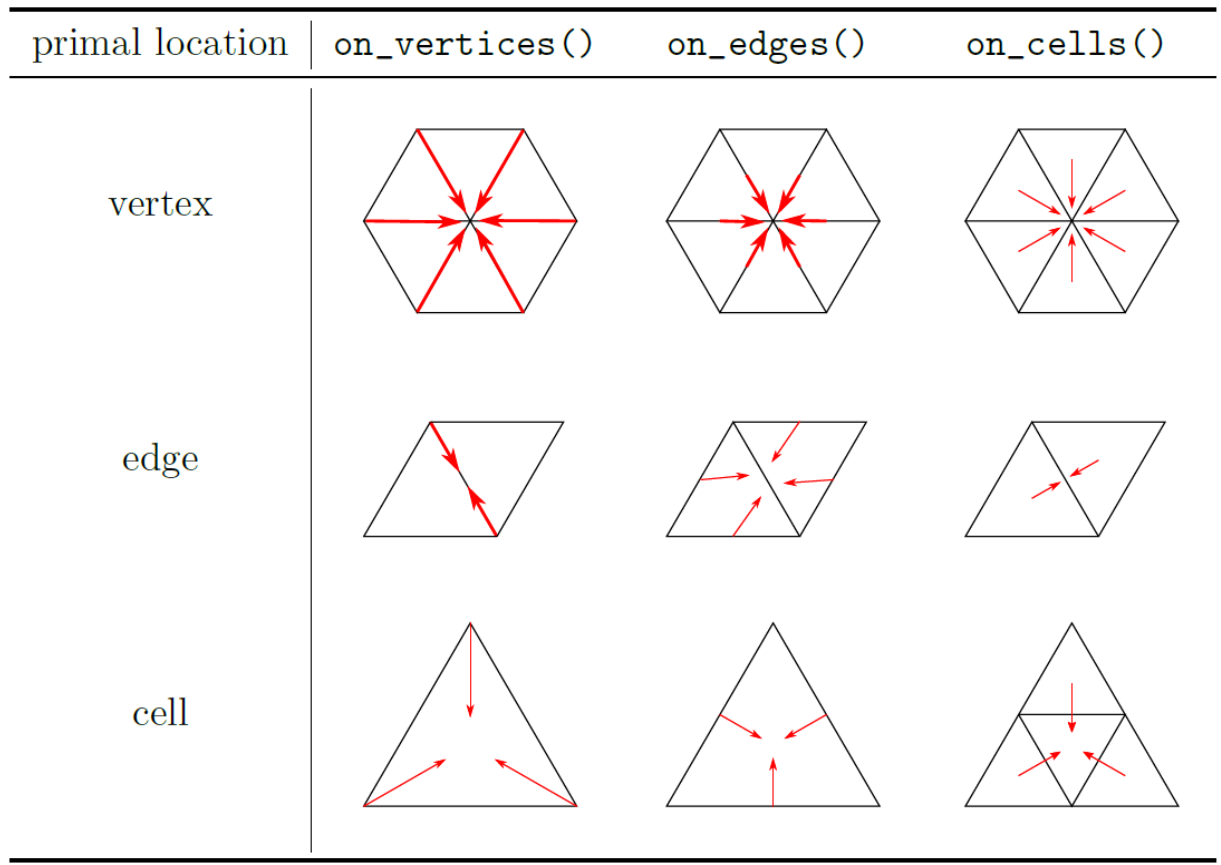

*Figure 5 All 9 types of stencil operations (accessing one level of neighbors) on irregular grids. Stencils are graphically depicted for triangular grids.*

Based on these basic stencil operations, the GridTools DSL was extended for irregular grids. An example of the syntax of the DSL is shown below for an MPDATA centered flux computation:

```
1   struct<uint_t Color> struct centred_flux {
2      using flux = accessor<0, out, edges>;
3      using pD = accessor<1, in, vertices>;
4      using vn = accessor<2, in, edges>;
5
6      template<typename Evaluation>
7      GT_FUNCTION static void Do(Evaluation& eval) {
8         eval(flux()) = 0.5*eval(vn()) * eval(on_vertices(sum, pD()));
9      }
10  };
```

Line 1 specifies specializations the operation for each color type (see Section 3.2).

Line 2-4 defines the parameters of the operator, similar to the syntax described in Section 3.1.2, with the extra specification of the location type.

Finally line 8 describes the stencil operation as a sum reduction over neighbor vertices. Since the location type of the output field (flux) is edges, the operator corresponds to the following reduction:

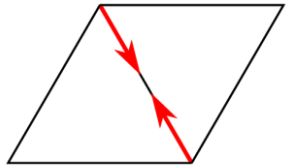

Therefore, the flux computation described in line 8 corresponds to a sum over the values of a two neighbor vertices of an edge (of the field being advected, pD), multiplied by the normal wind speed velocity to that edge.

The GridTools syntax allows describing any reduction operation over the basic stencil operators of Figure 5 by providing a custom functor:

```
0  auto reduction_sum =
     [](const double neighbour_value, const double accumulated_value) ->
double {
1      return neighbour_value + accumulated_value;
2  }
3  eval(flux()) = 0.5*eval(vn()) * eval(on_vertices(reduction_sum, 0.0,
pD()));
```

### 3.1.6 Access to connectivity information

The syntax elements introduced in Section 3.1.5 are general enough to implement basic operators on any irregular grid. However, it assumes that the operation performed over neighbors in the reduction function is totally symmetric. The syntax elements *on_cells, on_vertices, on_edges* cannot specialize the operation performed for each neighbor of the reduction.

A careful study of the methods employed in some of the dwarfs of ESCAPE, like MPDATA or the elliptic solver revealed in an early stage that this limitation is too restrictive and the DSL should give more control to specialize the operation for each neighbor.

An example of that requirement appears in a particular expression of the divergence operator in a triangular C-grid:

$$div(A)_i = \sum_{l \in \varepsilon(i)} v_{nl} {}^l\!/_{A_i}$$

Where i denotes a field location on cells, $\varepsilon(i)$ is the set of neighbor edges of a cell, $v_{nl}$ is the normal velocity at an edge, $l$ is the edge length and $A_i$ is the cell area.

A first implementation of this divergence operator can be achieved with the following GridTools syntax:

```
1   struct<uint_t Color> struct div {
2     using div = accessor<0, out, cells>;
3     using vn = accessor<1, in, edges>;
4     using edge_length = accessor<2, in, edges>;
5     using cell_area = accessor<3, in, cells>;
6
7     template<typename Evaluation>
8     GT_FUNCTION static void Do(Evaluation& eval) {
9       auto reduction_sum = [](double neigh_edgelength, double neigh_vn,
10             const double accumulated_value) -> double {
11         return neigh_edgelength * neigh_vn + accumulated_value;
12       }
13
14       eval(div()) = eval(on_edges(prod, edge_length(), vn())) /
                      eval(cell_area());
15   }
16  };
```

However there could be a more efficient implementation that avoids the division and the access to two fields $l$ and $A_i$, since the factor $l/A_i$ can be precomputed. The precomputed weight has to be stored as a multi-dimensional field located at *cell* location type, where the extra dimension has length 3, storing the ratio of edge length to cell area for each neighboring edge of a cell. Each iteration of the reduction would have to access a different element of the extra dimension of the weight, which is not supported by the *on_edges* syntax.

Another example of that limitation appears in the implementation of the upwind fluxes where the sign of the reduction operation depends on the sign of the velocity stored at that neighbor grid point.

In order to support these examples, GridTools introduces a new syntax element that gives full access to the connectivity information of the grid topology. The main functionality is provided by a connectivity class, which allows retrieving the tuple of offsets of any neighboring element for any location type.

The syntax

```
constexpr auto neighbour_offset = connectivity<vertices, cells, Color>::offsets();
```

returns a tuple of 6 elements, containing the offset of each neighbor cell of a node. That offset tuple is used for the reduction operation shown in the following figure

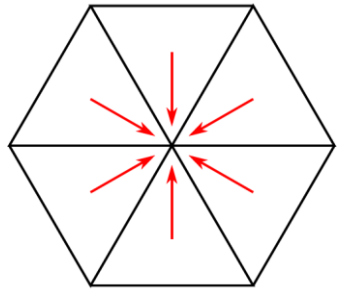

Although it provides a more verbose implementation as the equivalent *on_cells* syntax, it allows the user to specify custom reduction operations by using the retrieved offsets of the connectivity.

Using this functionality, the upwind operator can be implemented as follows:

```
0  template<uint_t Color> upwind_flux {
1    using flux = accessor<0, out, edges>;
2    using pD = accessor<1, in, vertices>;
3    using vn = accessor<2, in, edges>;
4
5    template<typename Evaluation>
6    GT_FUNCTION static void Do(Evaluation& eval) {
7      constexpr auto neighbour_offsets =
                connectivity<edges, vertices, Colo>::offsets();
8      constexpr auto ip0 = neighbour_offsets[0];
9      constexpr auto ip1 = neighbour_offsets[1];
10     float_type pos = eval(vn()) > 0 ? eval(vn()) : 0.;
11     float_type neg = eval(vn()) < 0 ? eval(vn()) : 0.;
12     eval(flux()) = pos * eval(pD(ip0)) + neg * eval(pD(ip1));
13   }
14 };
```

And the divergence example using a precomputed weight for $l/{A_i}$ can also be expressed using the connectivity functionality:

```
0  template<uint_t Color> div_precomputed {
1    using div = accessor<0, out, cells>;
2    using flux = accessor<1, in, edges>;
3    using weights = accessor<2, in, cells>;
4
5    template<typename Evaluation>
6    GT_FUNCTION static void Do(Evaluation& eval) {
7      constexpr auto neighbour_offsets =
                  connectivity<cells, edges, Colo>::offsets();
8      short e = 0;
9      double red{0.};
10     for(auto off : neighbour_offsets) {
11       red += eval(flux(off)) * eval(weight{edge(e)});
12       ++e;
13     }
14     eval(div()) = red;
15   }
16 };
```

### 3.2 Grid Indexing of global grids for efficient weather and climate operators

Memory data layouts of the fields where computations operate are crucial for performance, particularly on modern accelerators. Modern CPU processors operate on large vector widths (like AVX-512) while accelerators like NVIDIA GPUs compute on warps of 32 CUDA threads. Both provide more efficient use of memory if loads and stores are performed on aligned and coalescing accesses. A coalescing access will require that the memory loads/stores of different parallel cores of a GPU warp or elements of a vector instruction in a modern CPU processor are contiguous in memory.

Traditionally this was easy to achieve on lat-lon grid implementations, with memory layouts organized in rows/columns for the (i,j) indexing space.

On the contrary, memory layouts with good coalescing and alignment properties are much more challenging for irregular grids. For irregular grids without any regular pattern in the grid, it is not possible to obtain coalescing accesses. However, many grids employed by weather and climate models are derived from platonic solids, which retain their original structure. Examples are the octahedral grid (widely used in the ESCAPE dwarfs), the icosahedral grid or the cubed sphere.

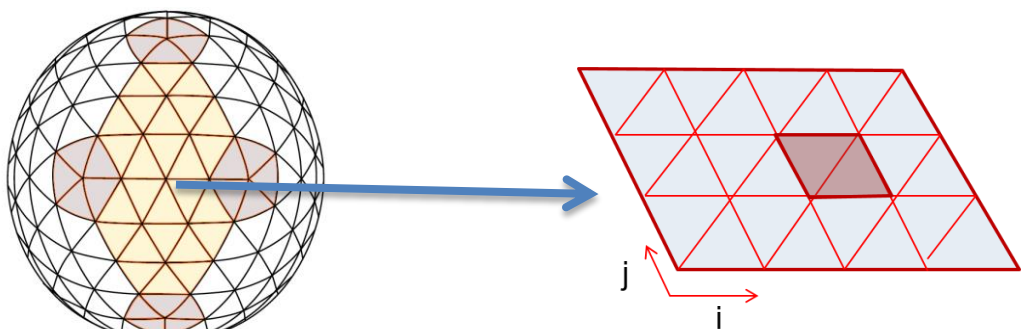

*Figure 6  Example of the structure in the icosahedral grid, present in each of the parallelograms*

Figure 6 shows the details of one of the parallelograms of a decomposition of the icosahedral grid. The figure on the right shows how the parallelogram can be indexed in a structured manner using rows, columns and color (upward/downward triangles), similar to lat-lon grids. However, the global grids present irregularities at each corner of the parallelograms of the original icosahedron.

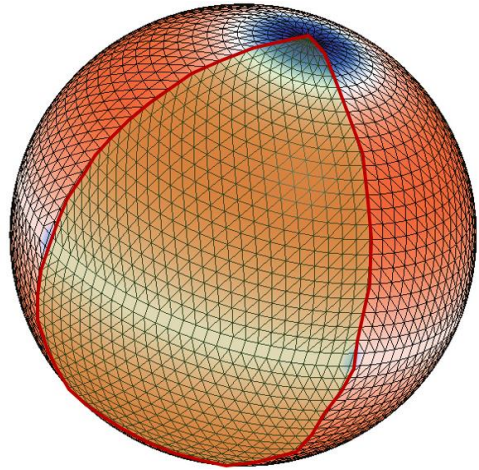

*Figure 7 Octahedral (O24) grid employed in several ESCAPE dwarfs*

Figure 7 shows the example of the O24 octahedral grid. Similar to the icosahedral grid, the octahedral grid can be decomposed in parallelograms that retain the original structure of the octahedron. Also the octahedral grid might contain irregularities, like the quadrilaterals (instead of triangles) at the equator and missing edges/nodes at the poles.

ESCAPE has explored how to obtain structured domain decompositions for the octahedral grids and conducted performance comparisons of different types of indexing on accelerators, with the goal to determine basic properties and optimal memory layouts for optimal performance of the ESCAPE dwarfs. Figure 8 (right) shows a possible domain decomposition of the octahedral grid that preserves the original structure of the octahedron.

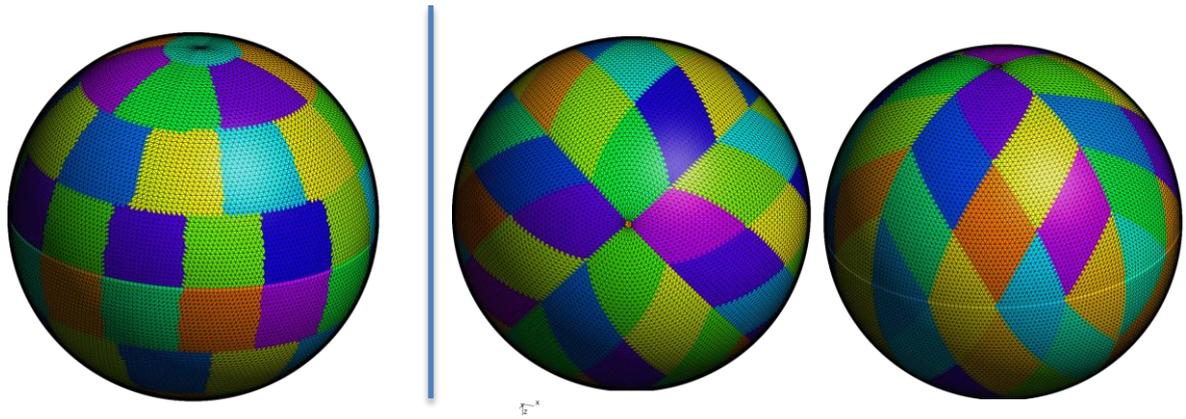

*Figure 8 Left: equal partitioning domain decomposition of the octahedral grid. Right: structured domain decomposition based on parallelograms*

With this domain decomposition, a new indexing method that uses coloring for cells/edges and vertices was introduced, with the important property that all the loads/stores performance by stencil computational patterns of the dwarfs are coalescing and a large fraction of them are also aligned.

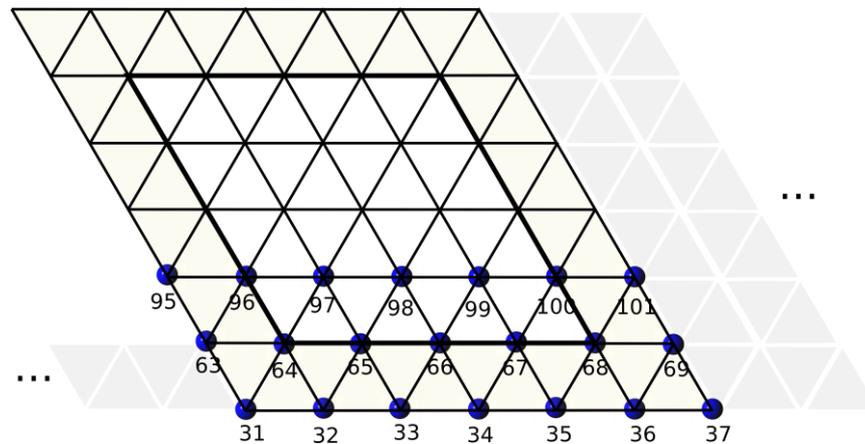

*Figure 9 Structured indexing of vertices of a parallelogram of the icosahedral/octahedral grid. Gray cells indicate padding inserted in order to align accesses to vertices within the compute domain. Number of colors for vertices is 1. Each vertex is indexed with a tuple (row, column)*

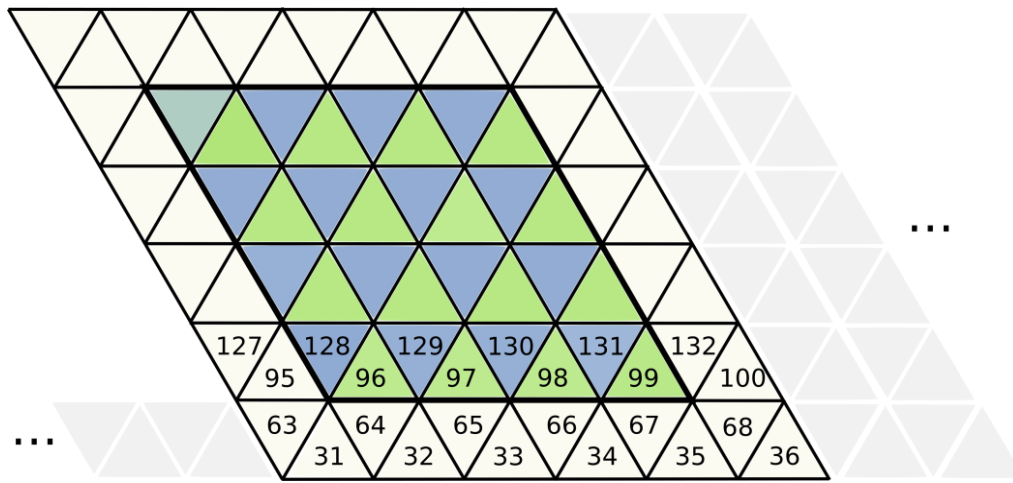

*Figure 10 Structured indexing of cells of a parallelogram of the icosahedral/octahedral grid (4x4 rows/columns and 1 row/column of halo). Gray cells indicate padding inserted in order to align accesses to cells within the compute domain. Number of colors for vertices is 2 (downward / upward triangles). Each cell is indexed with a tuple (row, color, column)*

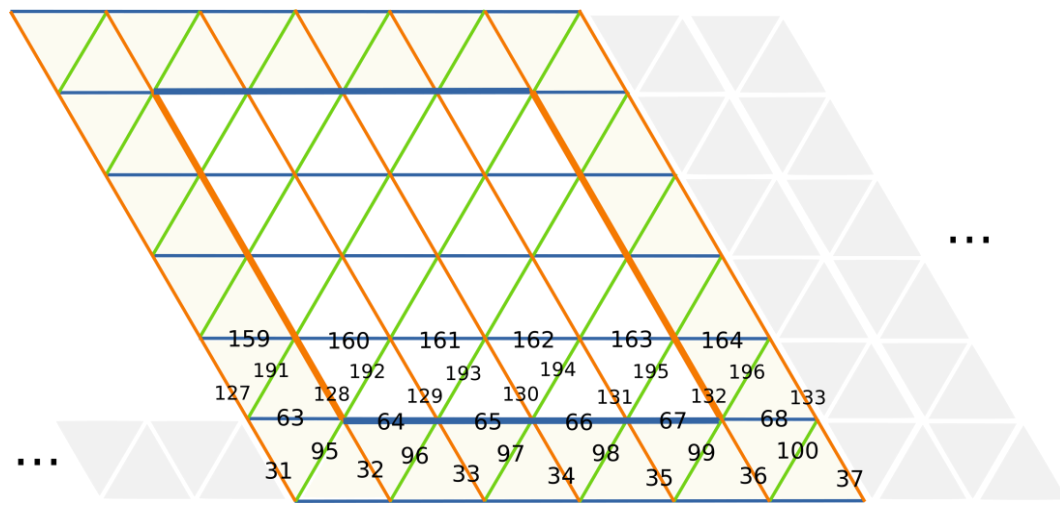

*Figure 11 Structured indexing of edges of a parallelogram of the icosahedral/octahedral grid (4x4 rows/columns and 1 row/column of halo). Gray cells indicate padding inserted in order to align accesses to edges within the compute domain. Number of colors for vertices is 1. Each edge is indexed with a tuple (row, color, column)*

Experiments on simple operators were conducted in order to evaluate different indexing and access methods for the ESCAPE dwarfs.

Three different indexing schemes were considered: a) the structured numbering (SN) proposed in Figure 9, Figure 10, and Figure 11, b) unstructured numbering (UN) similar to the default provided by Atlas for octahedral grids, Figure 12 and c) a Hilbert space numbering (HN) where the indexing of the cells follows a Hilbert space curve.

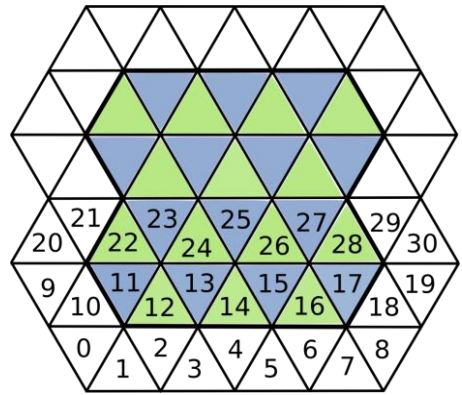

*Figure 12 Default indexing of Atlas, where the lack of color generates non coalescing loads/stores.*

Concerning the access method, direct access (DA) based on strides and indirect access using unstructured mesh like Atlas (IA) were studied.

The following table summarizes the results for a simple on_cells operation (row 3/column 3 of Figure 5) and a more complex operator that combines to on_cells reduction using an intermediate temporary field B.

*Table 1 Bandwidth (required data transfer / computation runtime) in GB/s of two stencil computations for the different indexing and access methods on an octahedral patch of size 128x128x80 for a P100 GPU.*

|       | $B = \sum_{neigh\_cells} A$ | $B = \left(\sum_{neigh\_cells} A\right) * fac1$ <br> $C = \left(\sum_{neigh\_cells} B\right) * fac2$ |
|-------|------|------|
| **SN DA** | 211 | 226 |
| **SN IA** | 270 | 269 |
| **UN IA** | 130 | 135 |
| **HN IA** | 256 | 240 |

Table 1 shows that the structured numbering yields best performance when combined with the indirect addressing. The DSL backend of GridTools for irregular grids developed within the ESCAPE project implements SN DA and UN IA. Future developments might support SN IA which gives best performance on GPU accelerators for grids that contain a structure and HN IA that provides still a good bandwidth for those grids for which a structure cannot be exploited.

## 4 Review of OpenMP and OpenACC programming models.

OpenMP and OpenACC are directive based approaches widely used for efficient parallelization of kernels on traditional CPUs as well as XeonPhi and NVIDIA GPUs, respectively. The use of directives for porting and accelerating the dwarfs of ESCAPE has been thoroughly described in D2.1 and D3.3. In this document we will use the work of these deliverables to evaluate and compare the OpenMP and OpenACC approaches with the DSL approach.

## 5 Implementation of ESCAPE dwarfs using a DSL and OpenACC

The DSL developments described in Section 3.1.3 have been used to implement a portable version of the MPDATA dwarf of the ESCAPE project. The MPDATA dwarf has been extensively described in the deliverable D1.2.

The MPDATA dwarf is implemented in FORTRAN as a sequence of subroutine calls, each one implementing the loops over edges/vertices and vertical levels with the corresponding computation on fields of each stage.

There are two types of computational patterns in the ESCAPE dwarfs using finite volumes: 1) **horizontal compact stencils** which require use of an unstructured mesh and 2) **vertical iterative solvers** derived from implicit solvers (like a tridiagonal solver).

Listing 4 and Listing 5 show the DSL implementation and Fortran OpenACC implementation of one of the horizontal stencil operators: the upwind flux computation. Both are similar in terms of structure and the code is equally readable. A first section will declare the field parameters. The indices of neighbor vertices of an edge are retrieved from the connectivity tables and used to compute the flux of pD.

The DSL version hides details away like the nested loops, the OpenACC directives used to specify properties of the GPU kernel and data layouts of the FORTRAN arrays.

```cpp
template<uint_t Color> upwind_flux {
  using flux = accessor<0, out, edges>;
  using pD = accessor<1, in, vertices>;
  using vn = accessor<2, in, edges>;

  template<typename Evaluation>
  GT_FUNCTION static void Do(Evaluation& eval) {
    constexpr auto neighbour_offsets =
            connectivity<edges, vertices, Colo>::offsets();
    constexpr auto ip0 = neighbour_offsets[0];
    constexpr auto ip1 = neighbour_offsets[1];
    float_type pos = eval(vn()) > 0 ? eval(vn()) : 0.;
    float_type neg = eval(vn()) < 0 ? eval(vn()) : 0.;
    eval(flux()) = pos * eval(pD(ip0)) + neg * eval(pD(ip1));
  }
};
```

*Listing 4 DSL implementation of the upwind flux computation of MPDATA*

```fortran
subroutine compute_upwind_flux(this, pflux, pD, pVn, iedge2node1, iedge2node2)
type(MPDATA_type), intent(inout) :: this
real(wp), intent(out) :: pflux(:,:)
real(wp), intent(in) :: pVn(:,:), pD(:,:)
real(wp) :: zpos, zneg
integer :: jedge, jlev, ip1, ip2
integer :: nb_edges, nb_levels
integer, intent(in) :: iedge2node1(:), iedge2node2(:)

nb_edges = this%geom%nb_edges
nb_levels = this%geom%nb_levels

!$acc parallel loop collapse(2) present(pflux, this, this%geom, this%geom%nb_edges, this%geom%nb_levels)
!$acc present(iedge2node1, iedge2node2)
!$acc copyin(nb_edges, nb_levels)
do jedge = 1, nb_edges
  do jlev = 1, nv_levels
    ip1 = iedge2node1(jedge)
    ip2 = iedge2node2(jedge)
    zpos = max(0._wp, pVn(jlev, jedge))
    zneg = min(0._wp, pVn(jlev, jedge))
    pflux(jlev, jedge) = pD(jlev, ip1)*zpos + pD(jlev, ip2)*zneg
  enddo
enddo
end subroutine compute_upwind_flux
```

*Listing 5 Fortran OpenACC implementation of the upwind flux computation of MPDATA*

```cpp
template<uint_t Color> struct upwind_fluz {
  using fluz = accessor<0, in, vertices>;
  using pD = accessor<1, in, vertices>;
  using wn = accessor<2, in, vertices>;

  template<typename Evaluation> static void Do(Evaluation& eval, kminimum) {
    float_type flz_p0 = eval(wn(k+1)*pD());
    float_type flz_p1 = eval(wn(k+1)*pD(k+1));
    float_type fluzbd = max(0.0, flz_p0)+min(0.0,flz_p1);
    eval(fluz()) = PIVBZ*fluzbd;

  }
  template<typename Evaluation> static void Do(Evaluation& eval, kbody) {
    float_type flz_pm1 = eval(wn()*pD(k-1));
    float_type flz_pc = eval(wn()*pD());
    float_type fluzbd = max(0.0, flz_pm1)+min(0.0,flz_pc);
    eval(fluz()) = PIVBZ*fluzbd;
  }
  template<typename Evaluation> static void Do(Evaluation& eval, kmaximum) {
    eval(fluz()) = PIVBZ * eval(fluz(k-1));
  }
};
```

*Listing 6 GridTools DSL implementation of the vertical upwind flux of MPDATA*

A comparison of the second type of computational pattern, i.e. the vertical implicit solver, can be found in Listing 6 and Listing 7.

```fortran
subroutine compute_upwind_fluz(this, pfluz, pD, pW, pivbz)

type(MPDATA_type), intent(inout) :: this
real(wp), intent(out) :: pfluz(:,:)
real(wp), intent(in) :: pW(:,:), pD(:,:), pivbz
integer :: jnode, jlev

!$ACC data present(pfluz, this, this%geom, this%geom%nb_nodes)
!$ACC data present(this%geom%nb_levels, pW, pD, pivbz)
!$ACC parallel loop collapse(2)
do jnode = 1,this%geom%nb_nodes
  do jlev = 2,this%geom%nb_levels
    pfluz(jlev,jnode)=max(0._wp, pW(jlev,jnode))*pD(jlev-1,jnode) &
      + min(0._wp, pW(jlev,jnode))*pD(jlev,jnode)
  enddo
  pfluz(1,jnode) = pivbz*pfluz(2,jnode)
  pfluz(this%geom%nb_levels+1,jnode) =
pivbz*pfluz(this%geom%nb_levels,jnode)
enddo
```

*Listing 7 Fortran OpenACC implementation of the vertical upwind flux of MPDATA*

The DSL makes use of specialized *Do* operators for each of the vertical regions as a way to specialize the computations at the boundary (*kmaximum* and *kminimum*). Again, both approaches are similar in terms of readability, with the particularity that the DSL abstracts away implementation details.

Furthermore, the DSL allows to compose multiple of these operators together, which is used by the library to apply advanced performance optimizations like loop fusion or software managed caches. Listing 8 shows the example of sequence of FORTRAN subroutine calls for a subset of operators of the MPDATA advection scheme. Listing 9 shows the equivalent code of the DSL, where the different operators are combined in a single kernel object.

```fortran
call compute_upwind_flux(this, zflux, pD, pVn,
     iedge2node1, iedge2node2)
call compute_upwind_fluz(this, zfluz, pD, pWn, rIVBZ)
call compute_fluxdiv(this, zdivVD, zflux, zfluz,
this%geom%dual_volumes, &
    inode2edges, inode2edges_size, inode2edges_sign)
call advance_solution(this, pD, zdivVD, prho)
```

*Listing 8 Sequence of operator calls for the subset of MPDATA in Fortran*

```cpp
m_upwind_fluxes = make_computation<gpu>(
    domain_uwf, grid_,
    make_multistage(execute<forward>(),
        cache(p_flux(), p_fluz(), p_divVD()),
        make_stage<upwind_flux, edges>(
                p_flux(), p_pD(), p_vn()),
        make_stage<upwind_fluz, vertices>(
                p_fluz(), p_pD(), p_wn()),
        make_stage<fluxzdiv, vertices>(
                p_divVD(), p_flux(), p_fluz(),
                p_dual_volumes(), p_edges_sign()),
        make_stage<advance_solution, vertices>(
                p_pD(), p_divVD(), p_rho())));
```

*Listing 9 Composition for the subset of operators of the MPDATA dwarf in the DSL language*

## 6 Comparison of different programming models for performance portability

In this section we will compare the different programming models used within the ESCAPE project to efficiently port dwarfs to various architectures and accelerators. The DSL implementation and its characteristics have been discussed in this document. The use of other programming models, i.e. OpenMP for CPU and XeonPhi and OpenACC for GPUs, and the performance obtained have been reported in deliverables D2-1 and D3-3.

### 6.1 DSL programming model evaluation

Since most of the weather and climate applications are memory bound on modern processors and accelerators, many of the performance optimizations focus on the best utilization of the memory subsystem of the computing architectures. Choices like memory layout of the fields and order of the loop nests are crucial to obtain good performance and they typically differ from one architecture to another. The abstraction of the implementation details provided by the DSL language allows to delegate those decisions to an architecture dependent backend.

In order to optimize memory bound kernels, one of the most prominent optimizations is the combination of tiling and loop fusion that increases data locality. All computing architectures offer a memory system with different levels of cache or scratch pad. Since the bandwidth of a cache level is typically orders of magnitude larger than main memory, the use of the cache of the memory system to reduce main memory accesses increases significantly the performance of the memory bound applications. Architectures like traditional CPUs or Intel XeonPhi have an automatic caching mechanism that does not require explicit instructions at the software level. However techniques like tiling or loop fusion are crucial in order to fit temporary computations into the fastest levels of the cache system.

On the other hand, NVIDIA GPUs require an explicit declaration in the programming model for the use of the different levels of the cache system, like the shared memory.

The composition of stages of the DSL (see Listing 9) allows the library to apply these loop tiling and fusion. In the MPDATA example shown for the computation of the fluxes, multiple fields are reused between the different computations. The theoretical calculations of main memory accesses with and without fusion are computed in Table 2.

*Table 2 Theoretical computation of the number of read and write accesses for the computation of the fluxes of MPDATA. The last column counts the total number of accesses per vertical plane for a grid wit 71424 nodes and 213199 edges. 2D fields are ignored*

|  | Edges access | Nodes access | Total |
|---|---|---|---|
| Without fusion | 1 r + 1 w | 7 r + 3 w | 1140638 |
| With fusion | 0 r + 0 w | 4 r + 1 w | 357120 |

More than a factor 2 in reduction of memory accesses to main memory is expected when the multiple operators of the fluxes are fused. Figure 13 shows the comparison of performance obtained on an Intel Haswell E5-2690 when fusion is applied.

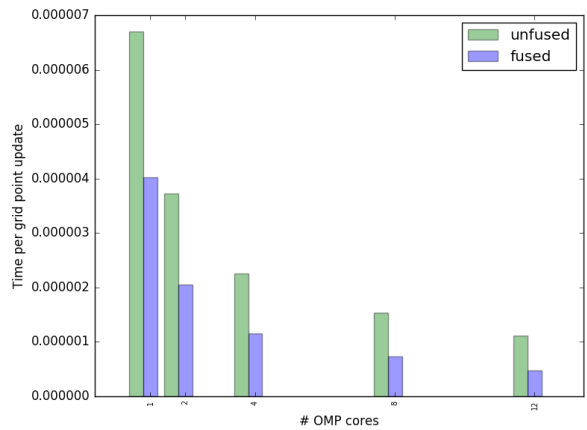

*Figure 13 Time per grid point update of the computation of the fluxes of MPDATA dwarf for domain of 128x128 nodes and 80 levels and for an Intel Haswell E5-2690 (gcc 5.3).*

Figure 14 shows the comparison between the OpenACC runs of the MPDATA scheme and the DSL kernels with the fusion and GPU specific optimizations on a K80. Even if recent OpenACC compilers show very competitive performance on modern GPUs, the use of domain specific information by GridTools library can outperform the performance obtained by the FORTRAN OpenACC kernel by a factor 2.1x. Among other optimizations, the use of a DSL allows to fuse all the stages that form a single computation (see Listing 10) of the MPDATA, using high bandwidth scratch pad for intermediate variable, which increases the data locality of the algorithm. Such optimizations can only be performed since the library assumes a parallel model that support only specific and limited computational patterns that can be expressed by the DSL language, as opposed to general purpose language compilers that cannot make such assumptions.

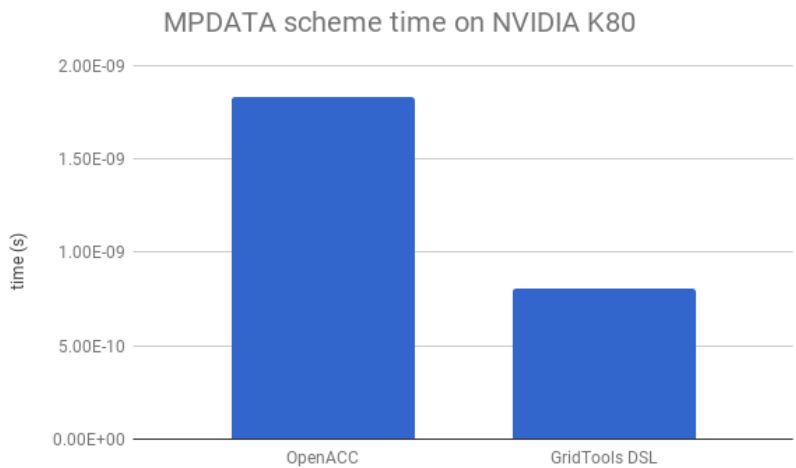

*Figure 14 Time of MPDATA scheme per grid point update (on a grid with 71424 nodes, 213199 edges and 80 levels) on a K80 NVIDIA GPU. OpenACC kernel are compiled with pgi 17.10 and GridTools with gcc 5.3 and CUDA 8.0*

## 6.2 Directives programming models evaluation

Deliverables D2-1 and D3-3 have reported on the use of directive programming models for portability of different dwarfs. The two approaches, OpenMP and OpenACC have in common the use of directives (comments in the host programming language) for inserting semantic about parallelization of the kernel and additionally on the GPUs to manage the separate memory space of the GPU.

An example of the use of OpenACC in the MPDATA dwarf has been introduced already in Listing 5. For comparison we show the equivalent optimized KNL kernel in Listing 10. A direct comparison between the two optimized implementations reveals the fact that often different computing architectures, depending on the dwarf or configuration, require different implementations like FORTRAN array data layouts or nest of the loops. This makes it impossible to retain a single source code that efficiently runs on multiple architectures. Depending on the dwarf choosing one of the implementation will show large performance deteriorations on other architectures.

```fortran
subroutine compute_upwind_flux(this, pflux, pD, pVn, iedge2node1, iedge2node2)
type(MPDATA_type), intent(inout) :: this
real(wp), intent(out) :: pflux(:,:)
real(wp), intent(in) :: pVn(:,:), pD(:,:)
real(wp) :: zpos, zneg
integer :: jedge, jlev, ip1, ip2

!$OMP PARALLEL PRIVATE(jedge, jlev, ip1, ip2, zpos, zneg)
do jlev = 1, nv_levels
  !$OMP DO SIMD SCHEDULE(STATIC, 256)
  do jedge = 1, this%geom%nb_edges
    ip1 = this%geom%iedge2node(jedge,1)
    ip2 = this%geom%iedge2node(jedge,2)
    zpos = max(0._wp, pVn(jedge, jlev))
    zneg = min(0._wp, pVn(jedge, jlev))
    pflux(jedge, jlev) = pD(jlev, ip1)*zpos + pD(jlev, ip2)
  enddo
  !$OMP END DO SIMD
enddo
!$OMP END PARALLEL
end subroutine compute_upwind_flux
```

*Listing 10 upwind flux computation of the MPDATA dwarfs optimized for XeonPhi*

Other than that, the task of parallelizing a single kernel to one computing architecture is relatively simple using the directive approaches.

Often not all the functionality of the host programming language, FORTRAN in this case, is supported by the OpenACC compiler. This forces the scientific developer to keep the numerical implementation in simple FORTRAN code, avoiding modern standard functionalities. An example of this is the use of deep copies for the GPU when using FORTRAN structures. This and other examples have been reported in deliverable D2.2.

The OpenAC standard contains a rich set of directives that can be used by FORTRAN codes to maximize the performance of the GPU kernels, like the use of the shared directive, which allows to use fast on-chip memory of the GPU. However, whether those keywords are used to efficiently to compile GPU code depends heavily on the compiler capabilities. Although compilers like pgi or cray have improved the performance delivered in recent versions, they are not as mature as cpu compilers. Therefore it can be expected that OpenACC compilers continue improving performance over time.

## 7   Conclusion

In this document we have presented a complete overview of the new developments and extensions to the GridTools DSL syntax in order to support methods of weather and climate models on irregular grids. This was the basis of an implementation of ESCAPE dwarfs using the DSL. We have presented the example of the MPDATA dwarf and compared in detail the implementations obtained by the DSL and the KNL (using OpenMP) and NVIDIA GPUs (OpenACC) ports, summarized in deliverables D2.1 and D3.3.

In the following we summarize some of the advantages and disadvantages extracted from the evaluation presented in this document.

OpenACC and OpenMP provide programming models which allow to relatively quickly port kernels of weather and climate models to accelerators like CPU and GPUs, since the implementations are based on directives that respect the original FORTRAN implementation of the model. Often it requires the transformation of loop nests and data layouts of the FORTRAN arrays. Additionally after a first port, a careful profiling of the application to investigate additional adaptations to optimize the kernels is required, as reported in D3.3.

For the use of the GridTools DSL a good level of expertise in the language is necessary, since it requires a rewrite of the model and a good understanding of the syntax elements of the DSL. The main disadvantages of the DSL are the need to rewrite the entirely of the numerical methods and the need of expertise to understand the language. The evaluations presented in this document show that both versions, DSL and explicit FORTRAN with OpenMP or OpenACC, retain similar level of readability. Also both require a certain level of boiler plate code, for example the definition of the parameters, which is equivalent in both cases.

Even if the DSL requires a good level of expertise, the advantage is that once the operators are implemented, the parallelization and efficient implementation for different architectures are not a responsibility of the user anymore, since it is delegated to specific computing architecture backends of the library.

The use of OpenMP and OpenACC often produces optimized versions of the code that differ for different architectures. Therefore, it is not possible to retain a single source code. The use of the DSL language instead, since it abstracts away the details of the architecture, makes it possible to retain a single implementation of the dwarfs that run efficiently for multiple architecture.

This document showed the importance of the indexing for obtaining good performance. The abstraction of the indexing in a completely irregular grid, when using an unstructured mesh, is done by Atlas. Therefore both approaches could equally benefit of optimizations of the indexing the maximizing performance for certain architecture. On the other hand, the use of a structured indexing, using row/color/columns for irregular grids with a structure like the octahedral or icosahedral is only possible with the use of the DSL.

Finally, the DSL provides a mechanism to compose multiple operators in a single kernel, which allows the library to apply data locality optimizations like loop tiling and loop fusion. As shown in the document, these optimizations make efficient use of the scratch pad and different cache levels of the chip, which reduces the amount of accesses to main memory. This allows the DSL implementations to obtain improved performance as compared to baseline directives implementations. These optimizations are only possible in combination with the structured indexing of the irregular grid and therefore they could not be obtained with the baseline FORTRAN implementations that make use of unstructured meshes provided by Atlas.

## Document History

| Version | Author(s) | Date | Changes |
|---|---|---|---|
| 0.1 | Carlos Osuna | 17th May 2018 | First version |
| 1.0 | Carlos Osuna | 5th June | Integrate reviews from M. Baldauf and P. Messmer |
| | | | |
| | | | |

## Internal Review History

| Internal Reviewers | Date | Comments |
|---|---|---|
| Michael Baldauf (DWD) | 31/05/2018 | Approved with comments |
| Peter Messmer (NVIDIA) | 30/05/2018 | Approved with comments |
| | | |
| | | |

## Effort Contributions per Partner

| Partner | Efforts |
|---|---|
| MSWISS | 5 |
| | |
| | |
| Total | 5 |

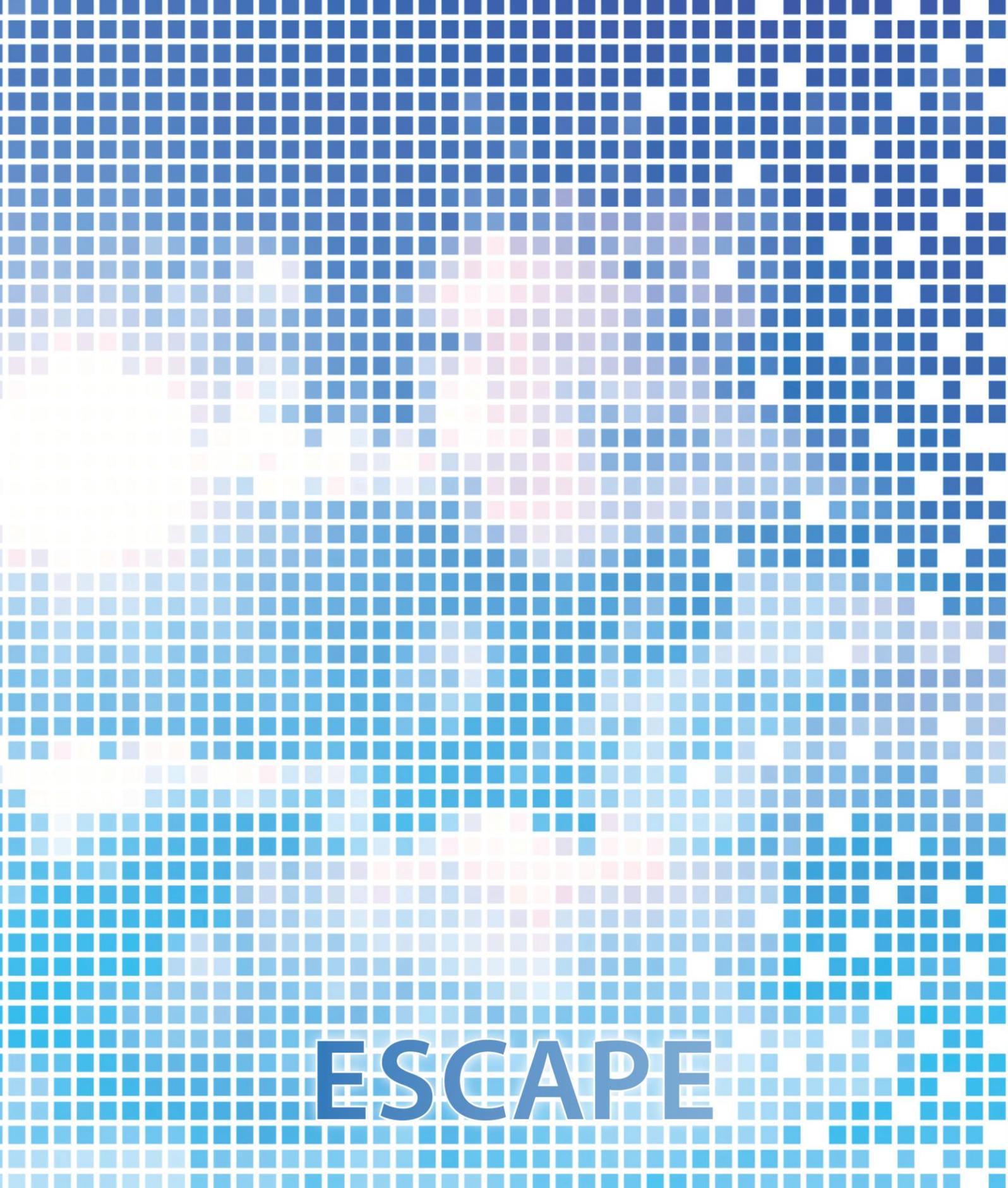